\documentclass[aip,CHA,reprint,amsmath,amssymb, floatfix, superscriptaddress, footinbib]{revtex4-1}
\usepackage{graphicx}
\usepackage{subfig}
\usepackage{placeins}
\usepackage{graphicx}
\usepackage{tabularx}
\usepackage{pgf,tikz}
\usepackage{placeins}
\usepackage[outdir=./]{epstopdf}
\usepackage{amsfonts, amsthm, mathrsfs}
\usepackage{listings}
\usepackage[english]{babel}
\usepackage{dcolumn}
\usepackage{bm}
\usepackage{natbib}
\usetikzlibrary{arrows}
\theoremstyle{plain}
\usepackage[outdir=./]{epstopdf}
\newcommand{\BigO}{\mathcal{O}}

\addto\captionsenglish{}
                                                                                          
\usepackage[bookmarks=false]{hyperref}

\begin{document}
\title{Macroscopic Models for Networks of Coupled Biological Oscillators}
\author{Kevin M. Hannay$^*$}
\affiliation{Department of Mathematics, University of Michigan, Ann Arbor, MI, 48109,  USA}
\author{Daniel B. Forger}
\affiliation{Department of Mathematics, University of Michigan, Ann Arbor, MI, 48109, USA}
\affiliation{Department of Computational Medicine and Bioinformatics, University of Michigan, Ann Arbor, MI, 48109 USA}
\author{Victoria Booth}
\affiliation{Department of Mathematics, University of Michigan, Ann Arbor, MI, 48109, USA}
\affiliation{Department of Anesthesiology, University of Michigan, Ann Arbor, MI, 48109, USA}
\date{\today}

\begin{abstract}
The study of synchronization in populations of coupled biological oscillators is fundamental to many areas of biology to include neuroscience, cardiac dynamics and circadian rhythms. Studying these systems may involve tracking the concentration of hundreds of variables in thousands of individual cells resulting in an extremely high-dimensional description of the system. However, for many of these systems the behaviors of interest occur on a collective or macroscopic scale. We define a new macroscopic reduction for networks of coupled oscillators motivated by an elegant structure we find in experimental measurements of circadian gene expression and several mathematical models for coupled biological oscillators. We characterize the emergence of this structure through a simple argument and demonstrate its applicability to stochastic and heterogeneous systems of coupled oscillators. Finally, we perform the macroscopic reduction for the heterogeneous stochastic Kuramoto equation and compare the low-dimensional macroscopic model with numerical results from the high-dimensional microscopic model. 
\end{abstract}

\maketitle

The study of coupled oscillators is important for many biological and physical systems, including neural networks, circadian rhythms  and power grids \citep{Winfree1980, Filatrella2008, strogatz2003sync}.  Mathematical models of these coupled oscillator systems can be extremely high-dimensional, having at least as many degrees of freedom as the number of oscillators. However, this microscale complexity is belied by the elegant simplicity which emerges at the macroscopic scale in many coupled oscillator systems. Quite generally, these systems demonstrate a phase transition as the coupling between the oscillators is strengthened leading to the emergence of a self-organized synchronized state \citep{Winfree2002}. 

This emergence of a synchronized state from the dynamics of a very high-dimensional dynamical system, suggests that a low dimensional representation of this system should be possible. A major step in this direction was proposed by Art Winfree in 1967 when he intuitively grasped that for systems of weakly coupled oscillators the time evolution of each limit cycle oscillator may be described by a single phase variable \citep{Winfree1967}. This method is now known as phase reduction and has been applied to study of many coupled oscillator systems \citep{Y.Kuram, Winfree1980, Schultheiss2011}. 

In the following years, Kuramoto formalized the mathematical procedure for phase reduction and used it to derive his now famous model for $N$ coupled heterogeneous oscillators,
\begin{align}
&\dot{\phi}_i=\omega_i+\frac{K}{N} \sum_{j=1}^N \sin(\phi_j-\phi_i),& i=1,N
\end{align}
where $\phi_i$ gives the phase of the $i$th oscillator, $K$ the coupling strength and $\omega_i$ gives the natural frequency of the oscillator \citep{Y.Kuram}. The natural frequencies of the oscillators are typically assumed to be drawn from some distribution $g(\omega)$ which reflects the heterogeneity in the oscillator population. The Kuramoto model captures the essential features of many coupled oscillator systems and has been used to study the phase transition to synchrony in detail \citep{Strogatz2000}. 

However, many biological systems contain thousands of oscillators, making even the phase model a very high-dimensional representation of the dynamical system. A recent breakthrough occurred when Ott and Antonsen discovered an ansatz that may be used for a family of Kuramoto-like systems to derive a low-dimensional model for the macroscopic behavior of the system \citep{Ott2008}. If the ansatz holds, the long-time behavior of a system of $N\rightarrow \infty$ oscillators can accurately be described by two differential equations, one for the mean phase of the coupled oscillators, and the other for their collective amplitude \citep{Ott2009}. Despite the hundreds of recent papers on the Ott-Antonsen (OA) ansatz, the authors are not aware of any carefully done experiments to test whether this powerful ansatz holds for biological systems. 

In this work we test the applicability of the Ott-Antonsen ansatz using a recent experimental data set collected from circadian oscillator neurons and through simulations of several models of coupled biological oscillators \citep{Abel2016}. We find the core assumptions which allow for the derivation of the macroscopic model using the OA ansatz are not valid in our test systems. However, we find a different, but related, ansatz is capable of describing the data well. Using a simple argument we are able to demonstrate the validity of our ansatz for a wide-class of models. Finally, we demonstrate how our ansatz may be used to derive macroscopic models for coupled oscillator systems.

\section*{Results}
The development of the Ott-Antonsen ansatz initiated a revolution in the coupled oscillator literature \citep{Pikovsky2015}. The method was first presented as an ansatz which could be used to reduce Kuramoto (and closely related systems) to a low-dimensional set of macroscopic equations \citep{Ott2008}. Remarkably, Ott and Antonsen were also able to show their procedure captures all the long-time attractors of these systems \citep{Ott2009}. The fact that this procedure allows for the derivation of strong analytic results has led to its application to a vast array of questions in the coupled oscillator community \citep{Luke2013, Hannay2015, Montbrio2015}. Recently, the Ott-Antonsen procedure was applied directly to the study of circadian rhythms for the first time \citep{Lu2016}.

While an extremely powerful tool, the Ott-Antonsen procedure suffers from several limitations. First, it may only be applied to systems which have a single harmonic in the coupling function describing the interaction between the oscillators \citep{Marvel2009}. Secondly, the ansatz is not valid for systems whose oscillators evolve with a stochastic component \citep{}. Each of these could severely limit its applicability to biological systems: Coupling between biological oscillators often features higher harmonic components in the coupling \citep{Hansel1993, Breakspear2010}, and the biological oscillators are invariably noisy \citep{Breakspear2010}. 

The final restriction on the Ott-Antonsen procedure is one of practicality rather than a formal mathematical restriction. In its most powerful form the Ott-Antonsen procedure requires the assumption that the distribution of natural frequencies be given by a rational function $g(\omega)=a(\omega)/b(\omega)$, which is typically taken to be a Cauchy (Lorentzian) distribution,
\begin{equation}
g(\omega)=\frac{\gamma}{\pi[(\omega-\omega_0)^2+\gamma^2)]},
\label{Eq:Cauchy}
\end{equation}
where $\omega_0$ is the median frequency and $\gamma$ controls the strength of the heterogeneity in the oscillator population. The Cauchy distribution decays slowly as $|\omega| \rightarrow \infty$ giving it ``fat-tails'', or a significant density of oscillators at extreme frequencies relative to the median $\omega_0$. Making the Cauchy assumption on the frequency distribution is a crucial step in achieving the dimension reduction to macroscopic variables. For more general frequency distributions, the OA procedure is still mathematically valid, although it produces an infinite set of integro-ordinary differential equations rather than a low-dimensional macroscopic model \citep{OmelChenko2012}. Let us refer to the Ott-Antonsen procedure with the additional assumption of a Cauchy distribution of frequencies as Cauchy Ott-Antonsen (COA). 

The COA ansatz takes a particularly simple form when written in terms of the Daido order parameters of the phase distribution  \citep{Daido1996,Daido1992,Daido1993}. The Daido order parameters are given by,
\begin{align}
Z_m(t) =R_m(t)e^{i \psi_m(t)}=\frac{1}{N} \sum_{j=1}^N e^{i m \phi_j(t)},
\end{align}
where $\phi_j$ are the phases of the oscillators and $R_m$ the phase coherences and $\psi_m$ the mean phases. Typically, only the first term is considered $Z_1=R_1e^{i \psi_1}$ and is known as the Kuramoto order parameter. Here $R_1$ measures the collective amplitude in the population of oscillators with $R_1 \approx 0$ indicating desynchrony and $R_1=1$ perfect synchrony. The COA ansatz then becomes a simple geometric relation between the Daido order parameters,
\begin{subequations}
\begin{align}
&Z_m=(Z_1)^m & \\
&R_m=R_1^m & &\psi_m=m\psi_1 & \text{ COA} 
\end{align}
\end{subequations}
For a phase distribution which is unimodal and symmetric about its mean phase we expect the mean phase relation $\psi_m=m\psi_1$ to hold generally. In this work we will restrict to considering cases where the phase distribution is approximately unimodal and symmetric. However, the prediction that $R_m=R_1^m$ is more subtle and its efficacy has not been evaluated for biological systems. 

\begin{figure*}
  \includegraphics{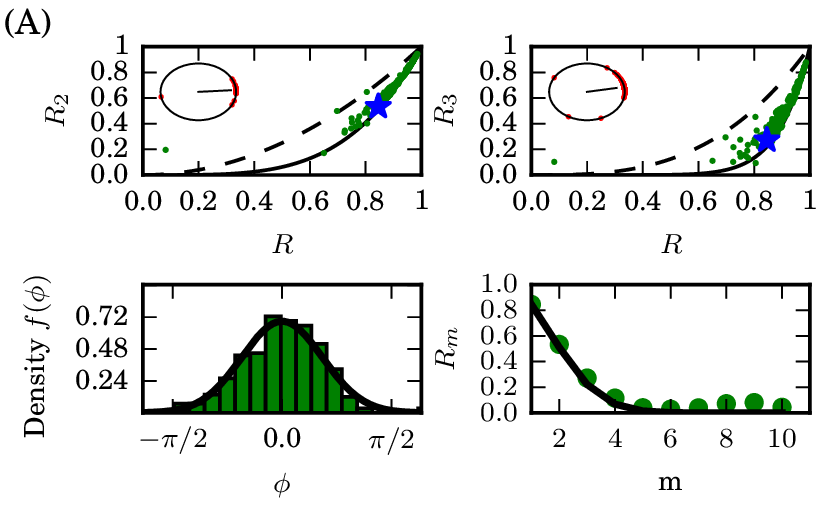}%
  \includegraphics{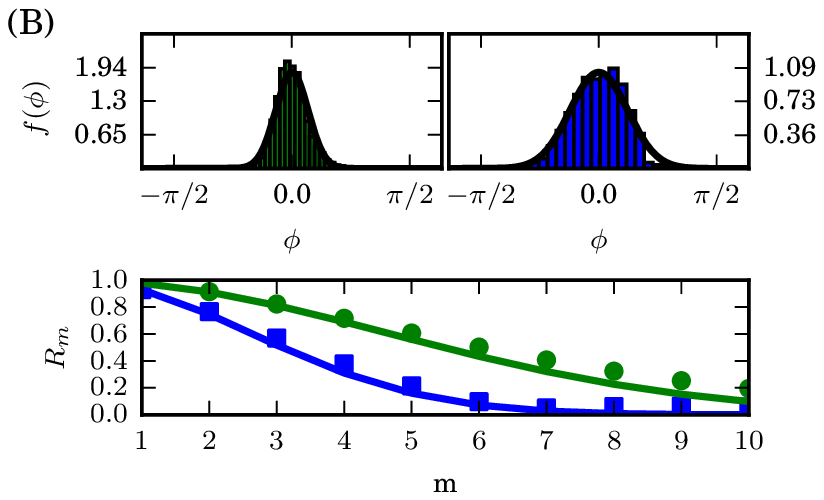}
  \includegraphics{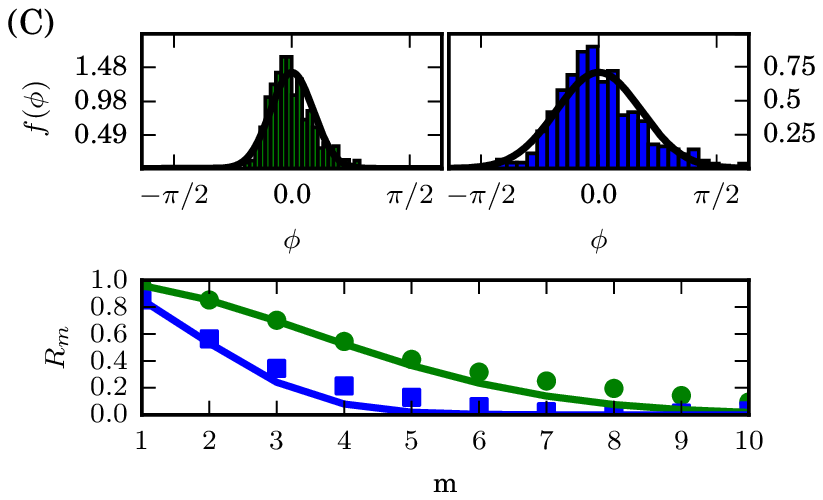}
  \includegraphics{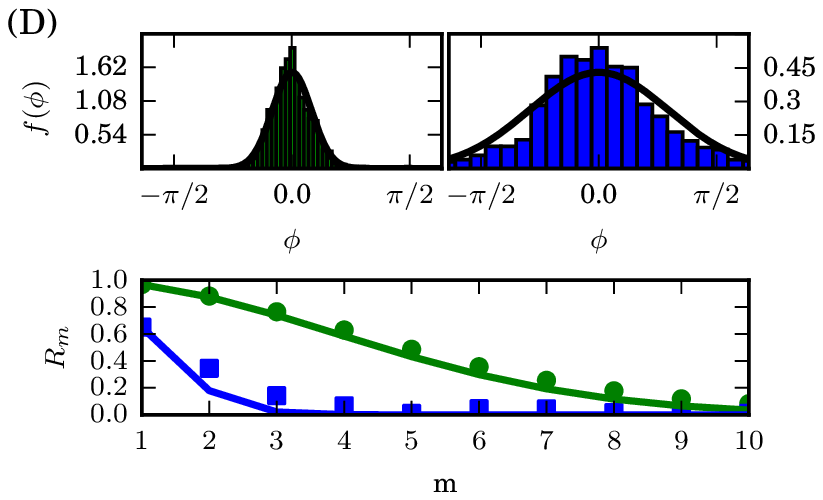}
\caption{A low-dimensional structure in the phase distribution of coupled oscillator systems. (A) Circadian oscillators data  \citep{Abel2016} (B) Simulation of a coupled heterogeneous Repressilator system \citep{Garcia-Ojalvo2004} (C) Simulation of a coupled heterogenous Morris-Lecar neuron system \citep{Morris1981} (D) Simulation of a coupled noisy modified Goodwin Oscillators model \citep{Kim2012}   (A) (top row)  Each green point shows a time measurement of the phase distribution of circadian oscillators. The solid black line shows the relation $R_m=R_1^{m^2}$ and the dashed line the COA relation $R_m=R_1^m$. Inset plots show $\psi_m-m\psi_1$. (A, bottom left) shows a histogram of the experimental phase distribution indicated by the blue star in in top row, against the $m^2$ ansatz phase distribution black line. (A,bottom right) We plot the first ten Daido order parameters for the experimental phase distribution (green dots) against the $m^2$ ansatz prediction. (B-D)  (top row) The simulated long-time phase distribution for the model as a histogram against the $m^2$ ansatz phase distribution for two different coupling strengths. (Bottom row) We plot the first ten Daido order parameters for the simulated phase distribution for two coupling strengths (green dots, blue squares) against the $m^2$ ansatz prediction.  }
\label{Fig:OADynGood}
\end{figure*}

To test this we computed the Daido order parameters for a recently published data set measuring the $\approx 24$ hour oscillations of protein expression in circadian neurons from whole suprachaismatic nucleus (SCN) explants \citep{Abel2016}. We examined this data set for evidence of the COA relation $R_m=R_1^m$ between the Daido order parameters Fig.~\ref{Fig:OADynGood}(A). However, we find this relationship is distinctly absent from the circadian data set. Additionally, numerical simulations of several other models of biological oscillators also reveal the COA approach provides a poor representation of the equilibrium phase distribution Fig.~\ref{Fig:OADynGood}(b-d). 

Instead we consistently find a different relation,
\begin{align}
&R_m=R_1^{m^2}  \text{,      } \psi_m=m\psi_1 & \text{        $m^2$ ansatz}
\end{align}
captures the properties of the phase distribution for these systems. What this means is that, while the Ott-Antonsen ansatz (COA) is a powerful tool for analyzing certain systems of coupled oscillators the diverse systems we test suggest a different scaling.  


 \subsection*{Emergence of the Scaling}

This alternate scaling may be derived under more general assumptions than those used by Ott and Antonsen. Let us consider the equilibrium phase distribution $\phi^*_j$ describing the states of the $N$ oscillators and assume $\phi^*_j \approx 0$ for each oscillator. Then a Taylor series expansion of the Daido order parameters may be written as,
\begin{align}
Z_m&\approx1+\frac{im}{N} \sum_{j=1}^N \phi_j^*-\frac{m^2}{2N} \sum_{j=1}^N (\phi_j^*)^2+... .
\end{align}
Then making use of our assumption that the equilibrium phase distribution is unimodal and symmetric we have that $\psi_m=m\psi_1$ and without loss of generality we may set $\psi_1=0$.  
Introducing the notation, $||\bm{\phi}^*||_k^k=\sum_{j=1}^N (\phi_j^*)^k$ gives,
\begin{subequations}
\begin{align}
&R_m \approx 1-\frac{m^2||\bm{\phi}^*||_2^2}{2N}\approx \left(    1-\frac{||\bm{\phi}^*||_2^2}{2N}\right)^{m^2}, \\
&R_m\approx R_1^{m^2}, \label{Eq:ansatz}
\end{align}
\end{subequations}
which will hold whenever the quantity $||\bm{\phi}^*||_2^2$ can be considered small. Which justifies the emergence of the $m^2$ ansatz we found in both the experimental and simulated data (Fig.~\ref{Fig:OADynGood}). 

\begin{figure}
\includegraphics{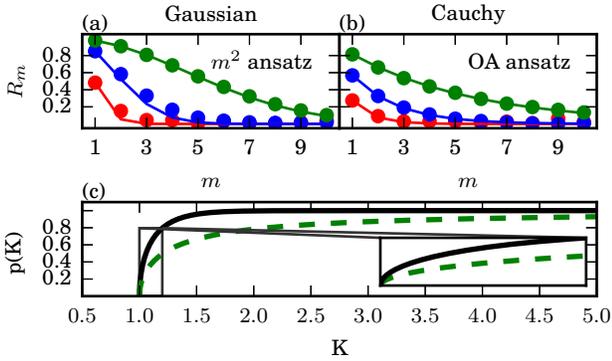}
\caption{(a,b) Circles show numerical results for the equilibrium phase distribution of the Kuramoto equation and curves show the ansatz predictions. Colors indicate the coupling strengths $K/K_c=1.1$ (red), $K/K_c=1.5$ (blue) and $K/K_c=3.0$ (green) (a) The $m^2$ ansatz for the Kuramoto equation with a Gaussian distribution of natural frequencies. (b) The COA ansatz  for a Cauchy distribution of natural frequencies (c) The fraction of locked oscillators ($p$) against the coupling strength $K$ for the Kuramoto model with a Cauchy distribution (dashed green) and Gaussian distribution (solid black) of natural frequencies. Parameters chosen such that $K_c=1.0$ for both distributions. \label{Fig:GaussCauchyP} }
\end{figure}

This analysis begs the question of how the COA relation $R_m=R_1^m$ and our relation can both be true. The root of the discrepancy is in the fat-tails of the Cauchy distribution. The slow decay in the tails of the Cauchy distribution property keeps the quantity $||\bm{\phi}^*||_2^2$ large for any finite coupling strength-as the oscillator population contains a significant fraction of oscillators which are not locked in a synchronized cluster. However, for natural frequency distributions with exponential tails (e.g. Gaussian) the fraction of locked oscillators grows quickly and we find our ansatz emerges for moderate coupling strengths. Our ansatz is compared with the COA ansatz in Fig.~\ref{Fig:GaussCauchyP}(a,b) for the Kuramoto system with a Gaussian and Cauchy distribution of frequencies. 

In fact, we may introduce a correction to our ansatz which takes into the account the fraction of oscillators which are phase-locked to the mean-field for a given coupling strength. Let $p$ be the fraction of the population which are locked to the mean-field, then we have $Z_m=pZ_m^{locked}+(1-p)Z_m^{drift}$ and $|Z_m^{drift}| \approx 0$ for the drifting population. Then the same Taylor-series based argument considering only the contribution of the locked population gives,
\begin{align}
&R_m\approx \frac{R_1^{m^2}}{p^{m^2-1}}, \label{Eq:AnWithP}
\end{align}
which collapses to the $m^2$ ansatz as $p \rightarrow 1$. Additionally, this analysis shows that assuming $p=1$ is expected to give a lower-bound on the Daido order parameter, in particular we have $R_m \geq R_1^{m^2}$ and $R_m \rightarrow R_1^{m^2}$ as $p \rightarrow 1$. For the Kuramoto model we may calculate $p(K)$ as,
\begin{align}
p(K)=\int_{-KR}^{KR} g(\omega) d\omega. \label{Eq:pFormula}
\end{align}
For the Kuramoto model with a $g(\omega)$ Gaussian or Cauchy we may solve for $p(K)$ using Eq.~\ref{Eq:pFormula}. The comparatively slow growth of the fraction of locked oscillators for the Cauchy distribution in shown in Fig.~\ref{Fig:GaussCauchyP}(c) relative to a Gaussian distribution of natural frequencies. 


\subsection*{Complex Networks and Noise}
The simplicity of our derivation makes it clear the $m^2$ ansatz should hold quite generally. In this section we characterize its convergence for the case of systems with complex network coupling and noisy oscillations. To explore this we consider a model for $N$ noisy heterogeneous phase oscillators, 
\begin{align}
\dot{\phi}_i=\omega_i + \frac{K}{d_i} \sum_{j=1}^N A_{ij} H(\phi_j-\phi_i)+\sqrt{D} \eta_i(t),
\label{Eq:GenPO}
\end{align}
where $\eta_i$ is a white noise process with $\langle \eta_i\rangle=0$ and $\langle\eta_i(t)\eta_j(t')\rangle=2\delta(t-t')\delta_{ij}$,
 $A_{ij}$ is an adjacency matrix and $d_i=\sum_{j=1}^N A_{ij}$ is the degree of the oscillator. Let $H$ be a $2\pi$ periodic coupling function and we assume that $H'(0)>0$. For simplicity we additionally assume that $A_{ij}$ defines a connected, undirected network. We note that Eq.~\ref{Eq:GenPO} is quite general and may be derived in many applications from higher dimensional limit cycle models under the assumption of weak coupling \citep{Arenas2008a}.

We consider the case of strong coupling between the oscillators such that, $\phi_j-\phi_i \approx 0$ for all oscillator pairs. In this case we may linearize about the phase locked state to give,
\begin{align}
\dot{\phi}_i =\tilde{\omega}_i -K H'(0)\sum_{j=1}^N L_{ij} \phi_j+\sqrt{D} \eta_i(t),
\label{Eq:GenPO1}
\end{align}
where $L$ is a normalized Laplacian matrix given by $L_{ij}=\delta_{ij}-A_{ij}/d_i$ and $\tilde{\omega_i}=\omega_i+KH(0)$. Our assumptions on the network mean that $L$ has real eigenvalues that may be ordered $\lambda_1=0\leq \lambda_2 \leq ... \lambda_N$ with associated eigenvectors $\{\bm{v}_1, ..., \bm{v}_N\}$ \citep{}. For this linear system we may solve for the quantity $\mathbb{E}\left[||\bm{\phi}^*||_2^2\right]_t$ using the Moore-Penrose pseudoinverse of the normalized Laplacian $L^\dagger$,
\begin{subequations}
\begin{align}
&\bm{\phi}^*= \frac{L^\dagger \bm{\tilde{\omega}}}{KH'(0)}, \text{ where \hspace{10pt}  } L^{\dagger}= \sum_{j=2}^N \frac{\bm{v}_j \bm{v}_j^T}{\lambda_j}, \\
&\mathbb{E}\left[ ||\bm{\phi}^*||_2^2\right]_t=\sum_{j=2}^N \left(\frac{|\bm{v}_j \cdot \tilde{\bm{\omega}}|}{\lambda_j KH'(0)}\right)^2+\frac{D}{\lambda_j KH'(0)}, \label{Eq:CNForm}
\end{align}
\end{subequations}
details of this derivation are given in the supplemental material. This analysis demonstrates that our ansatz will hold for sufficiently strong coupling strengths for any connected network where $||\bm{\tilde{{\omega}}}||$ is finite. Additionally, Eq.~\ref{Eq:CNForm} can be used to study how the speed of this convergence depends on the network connectivity, noise strength and the arrangement of the heterogeneous frequencies in the network \citep{Skardal2014}. 

These results are confirmed by numerical simulations of Eq.~\ref{Eq:GenPO} for the noisy and heterogeneous Kuramoto model ($H(\theta)=\sin(\theta)$) for various connectivity networks. In particular, we find the $m^2$ ansatz provides a quality approximation to the Daido order parameters for both Watts-Strogtaz small world \citep{Watts1998} and Barabasi-Albert scale-free \citep{Barabasi1999} network topologies. For each of these network topologies the accuracy of the approximation increases with the strength of the coupling as predicted by Eq.~\ref{Eq:CNForm}. 

\begin{figure}
\includegraphics{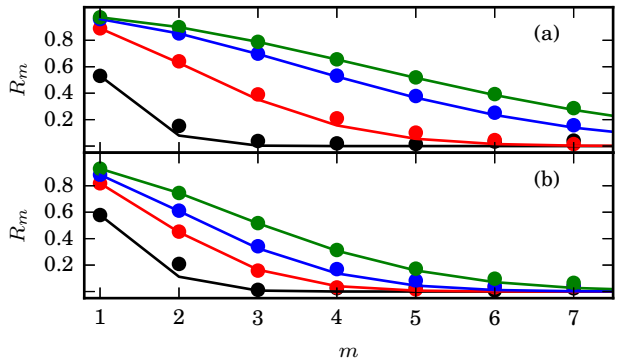}
\caption{The equilibrium phase distribution of complex network phase oscillators converges to the $m^2$ ansatz as the coupling strength between the oscillators increases. Circles show the results from simulations on networks of $N=1000$ coupled oscillators with noise strength $D=1$ and oscillator frequencies drawn from a Gaussian distribution with $\sigma=1$. Solid lines show $R_m=R_1^{m^2}$. Colors differentiate the coupling strengths  (a) Barabasi-Albert Scale-Free Network (b) Watts-Strogatz Small World Network \label{Fig:Networks}. Details of these simulations are given in the supplementary material. }
\end{figure}

\subsection*{Macroscopic Model}

A principal strength of the Ott-Antonsen approach is that the dynamics the Kuramoto model for a large system of coupled oscillators can be reduced to the following two variable differential equation:
\begin{subequations}
\label{Eq:OAMM}
\begin{align}
&\dot{R}_1=\left(\frac{K}{2}-\gamma\right) R_1 -\frac{K}{2}R_1^3 \\
&\dot{\psi_1}=\omega_0,
\end{align}
\end{subequations}
where $w_0$ is the median frequency of the oscillators and $\gamma$ is the dispersion parameter of the Cauchy distribution (Eq.~\ref{Eq:Cauchy}). The system provided by Eqs.~\ref{Eq:OAMM} provides a closed form model for the macroscopic properties of the Kuramoto system with a Cauchy distribution of natural frequencies. In this section, we demonstrate how the $m^2$ ansatz may be used to extract a similar macroscopic model for coupled biological oscillators. In particular we employ the $m^2$ ansatz as a motivated moment closure to extract a macroscopic model for the order parameter $Z_1$ for the noisy heterogeneous Kuramoto equation. That is we consider Eq.~\ref{Eq:GenPO} for a fully-connected network and $H(\theta)=\sin(\theta)$ with white noise such that $\langle \eta_i(t_1) \eta_j(t_2)\rangle=2 \delta(t_1-t_2)\delta_{ij}$. Under these conditions we may write our system using the Kuramoto order parameter $Z_1=R_1 e^{i \psi_1}$,
\begin{align}
&\dot{\phi}_i=\omega_i+K R_1\sin(\psi_1-\phi_i)+\sqrt{D}\eta_i(t). \label{Eq:IndMF}
\end{align}
If we consider the continuum limit by letting $N \rightarrow \infty$ in Eq.~\ref{Eq:IndMF} we find the continuity equation for the phase density function $f(\omega, \phi,t)$,
\begin{subequations}
\label{Eq:Cont}
\begin{align}
&\frac{\partial f}{\partial t}+ \frac{\partial}{\partial \phi}(f v )-D \frac{\partial^2 f}{\partial \phi^2}=0, \\
& v=\omega+K\Im[ e^{-i \phi}Z_1]],
\end{align} 
where $\Im$ denotes the imaginary part of the expression. 
\end{subequations}
We consider the Fourier series decomposition of $f$ given by,
\begin{align}
\label{Eq:OAansatz}
&f= \frac{g(\omega)}{2 \pi} \left\{1+\left[\sum_{n=1}^{\infty}A_n(\omega,t) e^{in \phi}+ \text{c.c.}  \right]    \right\},
\end{align}
where c.c. stands for the complex conjugate of the expression and $g(\omega)$ is the distribution of natural frequencies. Substitution of the Fourier series for $f$ into the continuity equation yields:
\begin{align}
&\frac{\dot{A}_{n}}{n}+(i \omega+Dn) A_n+\frac{K}{2} \left( Z_1 A_{n+1}-\bar{Z}_1 A_{n-1} \right)=0. 
\label{Eq:An}
\end{align} 
where barred quantities are the complex conjugate. In the continuum limit the Daido order parameters $Z_m$ are given by,
\begin{subequations}
\begin{align}
\label{Eq:ContOP}
Z_m(t)&=\int_0^{2\pi} \int_{-\infty}^{\infty} f(\omega, \phi,t) e^{i m \phi} d\omega d \phi  \in \mathbb{C} \\
&=\int_{-\infty}^{\infty} \bar{A}_m(\omega,t) g(\omega) d\omega, \label{Eq:CauchyInt}
\end{align}
\end{subequations}
using that all oscillating terms integrate to zero except $n=m$ in the Fourier series. If $g(\omega)$ is given by a Cauchy distribution (Eq.~\ref{Eq:Cauchy}) with median $\omega_0$ and dispersion parameter $\gamma$ we evaluate the integral in Eq.~\ref{Eq:CauchyInt} as a residue by arguing that $A_m(\omega,t)$ may be analytically continued into the complex $\omega$ plane \citep{Ott2008}.  Thus, for the Cauchy case we have that $Z_m(t)=\bar{A}_m(\omega_0-i\gamma,t)$. 
This substitution allows us to re-write Eq.~\ref{Eq:An} in terms of the Daido order parameters,
\begin{align}
\frac{\dot{Z}_n}{n}=(i \omega_0-\gamma-Dn) Z_n +\frac{K}{2} (Z_1 Z_{n-1}-\bar{Z}_1 Z_{n+1}). \label{Eq:MomentEq}
\end{align}
Finally, we set $n=1$ and apply the $m^2$ moment closure $Z_m=|Z_1|^{m^2-m} Z_1^m$ or $R_m=R_1^{m^2}, \psi_m=m\psi_1$, which yields an equation of motion for the Kuramoto order parameter $Z=Z_1$,
\begin{align}
&\dot{Z_1}=(i\omega_0-\gamma-D) Z_1 +\frac{K}{2}\left(Z_1-|Z_1|^2 (Z_1)^2 \bar{Z}_1\right)
\label{Eq:GenMF}
\end{align}
Separating the real and imaginary parts $Z=R_1e^{i \psi_1}$ gives,
\begin{subequations}
\label{Eq:HAFinal}
\begin{align}
&\dot{R}_1=\left(\frac{K}{2}-D-\gamma\right)R_1-\frac{K}{2}R_1^5\\
&\dot{\psi_1}=\omega_0.
\end{align}
\end{subequations}
In previous work Sonnenschein and Schimansky-Geier derived Eq.~\ref{Eq:HAFinal} for the special case of the stochastic Kuramoto model ($\gamma\rightarrow 0$) by employing an ad-hoc Gaussian moment closure on the phase distribution \citep{sonnenschein2013}. Interestingly, the Gaussian moment closure follows the $m^2$ ansatz found here. In accordance with our findings they found the macroscopic system (Eq.~\ref{Eq:HAFinal}) was able to capture the dynamics of the microscopic stochastic Kuramoto model accurately, particularly at strong coupling strengths. 

\begin{figure}
\includegraphics{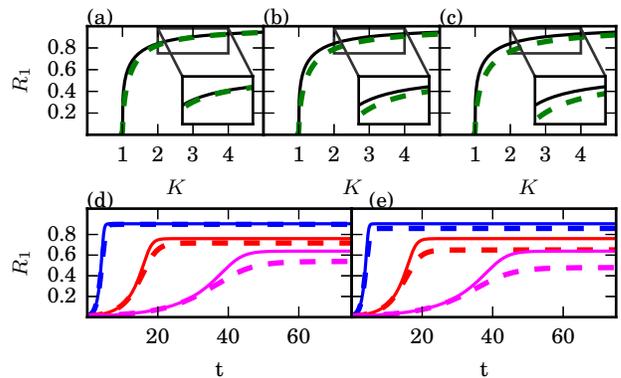}
\caption{(a-c) The equilibrium phase coherence $R_1$ against the coupling strength $K$ for the Kuramoto model for (a) $s=\gamma/D=0.05$ (b) $s=0.5$ (c) $s=1$. Solid black lines show the macroscopic model predictions and dashed (green) numerical simulations. Parameters chosen such that $K_c=1$ for the microscopic model. (d-e) The transient dynamics of $R_1$ for $K=1.2$ (magenta), $K=1.5$ (red) and $K=3.0$ (blue). The dashed lines show numerical simulations of the microscopic model and solid lines the macroscopic approximation (Eq.~\ref{Eq:HAFinal}). (d) s=0.05 (e) s=1.0.  \label{Fig:NHK}}
\end{figure}

Additionally, we find the $m^2$ ansatz provides an accurate approximation for the macroscopic dynamics of heterogeneous noisy Kuramoto model. In Fig.~\ref{Fig:NHK}, we show the predictions of the macroscopic model (Eq.~\ref{Eq:HAFinal}) against numerical simulations of the microscopic model (as estimated by using the first fifty moments of Eq.~\ref{Eq:MomentEq}\citep{sonnenschein2013}). 

In the limit of zero noise strength ($D \rightarrow 0$) the accuracy of the $m^2$ ansatz in seen to break down for Cauchy heterogeneity in the oscillators. This is to be expected given that the zero noise limit of Eq.~\ref{Eq:Cont} has been proven to follow the Cauchy Ott-Antonsen ansatz \citep{Ott2009}. However, in the case of weak to moderate heterogeneity relative to the noise strength $s=\gamma/D \leq 1$ we find the $m^2$ ansatz also provides an accurate description of the macroscopic dynamics (Fig.~\ref{Fig:NHK}). Moreover, we find the $m^2$ ansatz provides a useful upper-bound for the collective amplitude $R_1$ which tightens with increasing coupling strength. This may be explained by considering our result that $R_m\geq R_1^{m^2}$ and that $R_m \rightarrow R_1^{m^2}$ as the entire oscillator population is locked to the mean-field. 
 
As discussed the breakdown of the $m^2$ ansatz is related to the fat-tails in the Cauchy distribution, which cause the fraction of oscillators locked to the mean-field to grow slowly with the coupling strength. In most biological applications the heterogeneity in the population is unlikely to feature such extreme densities in the tails of the frequency distribution. This will only increase the accuracy of the $m^2$ ansatz in these cases.  In the next section we investigate how the $m^2$ ansatz may be used to derive macroscopic models for systems with strong heterogeneity.

\subsection*{Oscillator Heterogeneity}
In our macroscopic reduction of the noisy Kuramoto system we allowed for heterogeneity in the oscillators only by assuming a Cauchy distribution. However, our analysis has shown the $m^2$ ansatz is best applied to frequency distributions with exponential tails. For a general frequency distribution $g(\omega)$ the $m^2$ ansatz may be applied using,
\begin{subequations}
\begin{align}
&Z_1(t)=\int_{-\infty}^{\infty} \bar{A}_1(\omega,t) g(\omega) d\omega \\
& Z_m=|Z_1|^{m^2-m} Z_1^m. \label{Eq:mS}
\end{align}
\end{subequations}
However, without further simplification the advantage of our approach is largely negated as this inclusion results in an infinite set of integro-differential equations which only approximate the solution. A dimension reduction may be achieved when $g(\omega)$ takes the form of a rational function, which is usually taken to be the Cauchy distribution (Eq.~\ref{Eq:Cauchy}). In this case the $\omega$ dependence in the system collapses to the poles of the Cauchy distribution allowing for a macroscopic reduction, as seen in the derivation for the noisy Kuramoto equation. 

For a general symmetric and unimodal frequency distribution $g(\omega)$ with a maximum at $\omega_0$ we may think of approximating it with a Cauchy distribution $g_c(\omega, \gamma)$ which will allow for a reduction to a macroscopic model. Let $h(\omega,\gamma)=g(\omega)-g_c(\omega, \gamma)$ then we have,
\begin{subequations}
\begin{align}
&Z_1(t)=\bar{A}_1(\omega_0-i\gamma,t)+E_1(\gamma,t) \approx \bar{A}_1(\omega_0-i\gamma,t) \\
&E_1(\gamma,t)=\int_{-\infty}^{\infty} \bar{A}_1(\omega,t) h(\omega, \gamma) d\omega, \label{Eq:CauchyInt2}
\end{align}
\end{subequations}
Thus, the accuracy of the macroscopic reduction will depend on choosing the dispersion parameter $\gamma=\hat{\gamma}$, such that the magnitude of the error term $|E_1(\gamma,t)|$ is minimized. Additionally, the $m^2$ ansatz will give all the higher Daido order parameters with error $\BigO(E_1)$ using Eq.~\ref{Eq:mS}.  

The function $A_1(\omega,t) \in \mathbb{C}$ can be viewed as a frequency-dependent version of the Kuramoto order parameter $Z_1$. For oscillators which are entrained to the mean-field we may write,
\begin{align}
A_1(\omega,t)=\rho(\omega) e^{i(\theta(\omega)+\Omega t)},
\end{align}
where $\Omega$ gives the frequency of the mean-field, $\rho(\omega)$ describes the collective amplitude and $\theta(\omega)$ the entrainment angle for oscillators with natural frequency $\omega$. When oscillators with frequency $\omega$ are locked to the mean-field we have $\rho(\omega)=1$ \citep{OmelChenko2012}. 

For the Kuramoto model, oscillators with $\omega \leq KR$ are locked to the mean-field with $\theta(\omega)=\arcsin(\omega/KR)\approx \omega/KR$. Therefore we may approximate the magnitude of the error integral by considering only the locked oscillators,
\begin{align}
|E_1(\gamma)|\approx |L_1(\gamma)|=\left|\int_{-KR}^{KR} e^{i \frac{\omega}{KR}} h(\omega, \gamma) d\omega \right |. \label{Eq:MinGamma}
\end{align}
Thus, we solve for $\hat{\gamma}$ such that $|L_1(\gamma)|$ is minimized. For the frequency distributions we consider it is possible to find $\hat{\gamma}$ such that $|L_1(\hat{\gamma})|=0$. In general, $\hat{\gamma}$ will depend on the coupling strength $K$ both directly and implicitly through $R_1(K)$.

\begin{figure}
\includegraphics{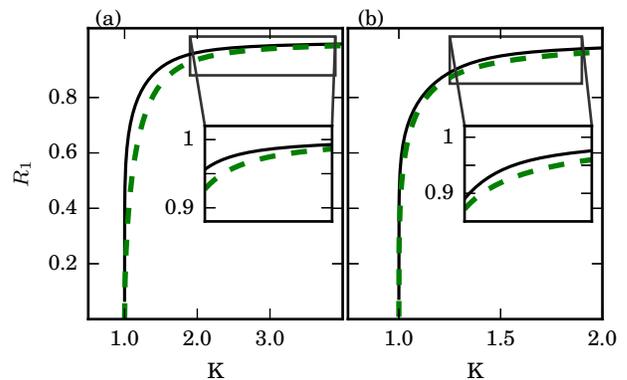}
\caption{The equilibrium phase coherence $R_1$ against the coupling strength $K$ for the Kuramoto model for (a) Gaussian (b) $\propto e^{-\omega^4/a}$ distributions of natural frequencies. Exact solutions are shown as dashed green, solution according to the $m^2$ ansatz solid black \label{Fig:KRCurves}}
\end{figure}

Therefore, for the heterogeneous Kuramoto model we have the macroscopic model for $Z_1=R_1e^{i\psi}$,
\begin{subequations}
\label{Eq:HMM}
\begin{align}
&\dot{R}_1=\left(\frac{K}{2}-\hat{\gamma}(K)\right)R_1-\frac{K}{2}R_1^5\\
&\dot{\psi}=\omega_0.
\end{align}
\end{subequations}
For $R K\approx 0$ we may solve for $\hat{\gamma}$ by setting $|h(\omega_0,\hat{\gamma})|=0$ which yields $\hat{\gamma}=1/[\pi g(\omega_0)]$. Therefore, the macroscopic model captures the critical coupling strength $K_c=2\hat{\gamma}$ as determined the classical self-consistency approach \citep{Strogatz2000, Y.Kuram}. Moreover, we find the macroscopic model (Eq.~\ref{Eq:HMM}) provides a close approximation to $R_1(K)$ as the coupling strength increases as shown in Fig.~\ref{Fig:KRCurves} for $g(\omega)$ Gaussian and $g(\omega) \propto e^{-\omega^4/a}$.

Finally, we note that the error in approximation of the integral (Eq.~\ref{Eq:MinGamma}) scales with the fraction of locked oscillators $p$. Thus, the the Cauchy approximation and the $m^2$ ansatz each introduce errors which scale with the fraction of locked oscillators. Therefore,  employing the Cauchy ansatz alongside the $m^2$ ansatz does not add any additional assumptions to the approximation and does little to effect the accuracy of the approach (Fig.~\ref{Fig:KRCurves}).

 \section*{Discussion}
 
 In this work we have reported a low-dimensional relationship which emerges between the Daido order parameters of the phase distribution functions of coupled oscillators. We have identified this relationship in \textit{in vivo} recordings of circadian oscillators and \textit{in silco} simulations of several models of biological oscillators. We demonstrate this relationship robustly emerges in networks of noisy heterogeneous coupled oscillators for sufficiently strong coupling strengths. 
 
 This analysis reveals that as the coupling strength increases in these systems the entire phase distribution of oscillators may be described through knowledge of the macroscopic Kuramoto order parameter $Z_1$- with all higher order Daido order parameters being slaved to the first order term. This result by itself could have many applications within neuroscience and mathematical biology. For instance, the ability to construct the entire phase distribution from knowledge of the collective amplitude $R_1$ should allow for an improved understanding of the effect of amplitude on phase-resetting in biological oscillators \citep{Hannay2015, Abraham2010}. This could result in an improved understanding of the entrainment of networks of coupled oscillators\citep{Abraham2010}. 
 
 Further, we have demonstrated this relationship may be used as a ``motivated moment closure'' to extract a low-dimensional model for coupled noisy heterogeneous oscillators. We extracted a mean-field model for noisy heterogeneous Kuramoto oscillators. Future work could generalize this procedure to allow for higher harmonics in the coupling function facilitating the analytical study of many models of biological oscillators. 
 
 A principal strength of the motivated moment closure approach is that the parameters and variables of the derived macroscopic model have direct physical interpretations. Therefore, the predictions of the model may be compared with experimental data from the cellular, tissue and whole organism levels. This feature should allow macroscopic models to be derived in many applications which may be compared with experimental data to provide fundamental insights into the functioning of coupled oscillator networks \citep{Lu2016}. 

The low-dimensional system we derive differs slightly from the Ott-Antonsen approach as it produces a term of order $R^5$ in the amplitude equation as compared with the cubic scaling $R^3$ in the Ott-Antonsen equations \citep{Ott2008, Lu2016}. We note that a cubic scaling is expected for coupling strengths near the critical coupling strength $K_c$ as the normal form for a Hopf bifurcation \citep{Guckenheimer2013}. Therefore, we expect our ansatz to provide an overestimate of the growth of the phase coherence about the critical coupling strength and conclude is not an appropriate tool for studying the scaling of the order parameter about the critical coupling. However, we find our approach converges to the correct value as the coupling strength increases. Additionally, we find the rate of this convergence is determined by the fraction of oscillators which are locked in a synchronized cluster. 

Moreover, we note that higher-order terms in the amplitude growth have previously been required to accurately model the collective amplitude dynamics of the human circadian rhythm in response to a desynchronizing light-pulse \citep{Jewett1998}. The $R^5$ term predicts it should be difficult to to increase the amplitude of the circadian rhythm by applying light pulses to an equilibrium circadian amplitude. This is in accordance with experimental results that light pulses administered during the day do not significantly effect the circadian amplitude \citep{Jewett1994b, Jewett1997a}. Finally, we note that a previous comparison between two phenomenological van der Pol models for human circadian rhythms showed the model with higher order terms better explained human circadian amplitude data \citep{Indic2005a}.

 \section*{Methods}
 
The circadian time-series shown in Fig.~\ref{Fig:OADynGood}(a) was collected as described in Abel et al \citep{Abel2016}, who generously made their data set publicly available.  Briefly, the time-series was collected from whole SCN mouse explants cultured for ~14 days. The expression of the circadian marker PERIOD2::Luciferase was monitored under a microscope, with bioluminescence measurements collected every hour. On day six in culture tetrotoxin (TTX) was added to the culture in order to block neuronal signaling and desynchronize the neurons. The TTX solution was washed away and the culture was allowed to resynchronize. For our purposes we removed the time-points when the TTX solution was added in order to study the phase distribution of the coupled clock neurons. 

The raw bioluminscience data were processed following established methods \citep{Myung2012}. First,  the raw bioluminscience data was de-trended by removing the Hodrick-Prescott (H–P) baseline trend with a large penalty parameter $\lambda=10^6$ to minimize loss of the oscillatory signal component. The time-dependent protophase of each oscillator was extracted by dimensional embedding with an eighteen hour embedding lag \citep{Small2005}. Finally, the time-dependent phase was estimated using the protophase to phase transformation as specified in the DAMOCO Matlab toolbox \citep{Kralemann2007, Kralemann2008}. 

Details for the mathematical models used in Fig.~\ref{Fig:OADynGood}(b-d) are given in the supplementary material. The estimation of the phase distribution for the the \textit{in silco} data was carried out in much the same manner as described for the experimental data. However, due to the large number of data points available in the simulated data we used the Hilbert transform to estimate the protophase of the oscillators. 

\begin{acknowledgments}
The authors would like to thank S. Strogatz, J. Myung, J. Abel,  A. Stinchcombe and A. Cochran for useful discussions related to this work. This research was supported by Human Frontiers of Science Program Grant RPG 24/2012 and National Science Foundation Grant DMS-1412119.
\end{acknowledgments}

 \bibliographystyle{naturemag_noURL}

\end{document}